\documentclass[aps,twocolumn]{revtex4}%
\usepackage[colorlinks,linkcolor=blue,citecolor=blue,hyperindex,bookmarks=false,pdfstartview=FitH]{hyperref}
\usepackage{gensymb}
\usepackage{amsfonts}
\usepackage{amsmath}
\usepackage{amssymb}
\usepackage{epsfig}
\usepackage{graphicx}
\usepackage{color}

\providecommand{\U}[1]{\protect\rule{.1in}{.1in}}
\providecommand{\U}[1]{\protect\rule{.1in}{.1in}}
\providecommand{\U}[1]{\protect\rule{.1in}{.1in}}
\providecommand{\U}[1]{\protect\rule{.1in}{.1in}}

\providecommand{\U}[1]{\protect\rule{.1in}{.1in}}

\begin{document}
\title{Chiral phase modulation and tunable broadband perfect absorber using the coherent cold atomic ensemble}
\author{Yi-Xin Wang$^{1}$, Yan Zhang$^{1, \ast}$, Lei Du$^{2}$ and Jin-Hui Wu$^{1}$}
\affiliation{$^{1}$ School of Physics and Center for Quantum Sciences, Northeast Normal University, Changchun 130024, China}
\affiliation{$^{2}$ Department of Microtechnology and Nanoscience (MC2), Chalmers University of Technology, Gothenburg S-412 96, Sweden}
\affiliation{$^{\ast}$zhangy345@nenu.edu.cn}
\date{\today }


\begin{abstract}
We investigate the two-channel nonreciprocal scattering of a coherent atomic ensemble under the linear spatial Kramers-Kronig modulation, which has potential applications in chiral phase modulation and broadband coherent perfect/asymmetric absorber that yet is typically unavailable in conventional continuous atomic media. In the regime of electromagnetically induced transparency, we observe the direction-dependent (chiral) phase modulation, which may enrich the burgeoning chiral quantum optics and can be used for implementing photonic filters, unidirectional amplifiers, and coherent asymmetric absorbers. By simplifying the stringent generation condition of coherent perfect absorption (CPA), we demonstrate the possibility of realizing two-channel CPA with broadband and sharp edges. Our proposal may be used to design and integrate some all-optical functional devices at extremely low power levels for quantum information processing and optical communication networks.
\end{abstract}
\maketitle

\section{Introduction}
The ability of asymmetric (chiral) manipulation on photon flows is a crucial technology for both classic and quantum communications, enabling nonreciprocal transmission and amplification \cite{unT1,unT2,unT3,unT4,unT5,unT5a,unT6,unT7}, unidirectional reflection \cite{uRmov1, KK1}, chiral topological states \cite{uTS1,uTS2}, nonreciprocal chaos \cite{unC1,unC2}, and nonreciprocal quantum entanglement and steering \cite{uES1,uES2}, to name a few. This thus inspires growing research interest in corresponding photonic circuits and devices \cite{dev1,dev2,dev3}. Recently, there has been a focus on phase nonreciprocity and its applications, such as the nonreciprocal phase shift \cite{nonP1}, phase transition \cite{nonP2}, and phase modulation \cite{nonP3}. Furthermore, the nonreciprocal phase sensitivity enables asymmetric transmission of waves, which plays a crucial role in specific information communication schemes involving one-way signal-controlling tasks \cite{unT9}. 

However, light transmissions are typically reciprocal for normal natural and artificial media as per the Lorentz reciprocal theorem \cite{lorentz1}. Moreover, the atomic media also usually exhibit reciprocal light reflection, and the reflection can be made perfect by utilizing photonic bandgaps as a fundamental photonic manipulation technique \cite{PBG1, EPBG1, PBGa3, PBGa4, EPBG3, PBGa5, PBG2}. Given this, a large number of efforts have been made to break the reciprocity of these systems, with the underlying physical mechanisms including magneto-optical effect \cite{MOF1}, Doppler effect \cite{unT1, unT4}, moving atomic lattice \cite{unT3,uRmov1,uRmov2}, optical parity-time symmetry \cite{uPT1,uPT2,uPT3}, and so on. Recently, the spatial Kramers-Kronig (KK) relation has opened up a promising route for developing nonreciprocal optics \cite{KK1}, which can physically and mathematically explain many unidirectional reflectionless behaviors based on non-Hermitian optical modulations without requiring partial or time-reversal symmetries of the refractive index. 
However, engineering specific spatial structures of media remains challenging, and previous schemes with fixed spatial KK media lack the tunability \cite{KK2, KK3, KK4, KK5}. 
Given this, a more advanced technique called spatial KK modulation has been developed. More specifically, one uses a laser beam with a linearly spatial-varying intensity to drive atoms, thus modulating the complex reflective index to satisfy the spatial KK relation \cite{KK6}. 


As an excellent candidate for controlling light with light \cite{CPAnon1, CPAnon2}, coherent perfect absorbers have been demonstrated in systems where the incident energy is delivered through multiple channels, such as the single-channel \cite{CPAsingle} scattering scheme and the two-channel scattering scheme that involves a medium being bilaterally illuminated with a complex potential. The underlying phenomenon, coherent perfect absorption (CPA) arising from the \emph{interplay of interference and absorption} \cite{CPArevline, CPA01}, can be interpreted as the time-reversed lasing or anti-lasing effect \cite{CPAla1, CPAla2}, and its implementation does not require time-reversal symmetry \cite{CPAla4, CPAla5}. The nonlinear response of CPA to the incident waveform allows for flexible control over light scattering and absorption \cite{CPAnon2, ctrlig}. 
In data processing, CPA provides a means of modulating a signal by another coherent optical signal under a linear strategy, different from that used with nonlinear optical media \cite{nonopt2, nonopt3}. This enables the implementation of linear optical switches, modulators, and logic gates \cite{CPAnon1,ctrlig, CPA02, CPAlog}, which can be operated at extremely low power levels. 
The perfect absorber has been demonstrated in various systems, such as the metasurface/metaatoms \cite{CPA03}, cavity system \cite{CPAcav1, CPAcav2, CPAcav3}, chiral medium \cite{CPA04}, and optical lattice \cite{ucpa}. 
With the realizations with quantum optics, CPA has found applications in the quantum fields, particularly in scenarios involving few-photon and entangled-photon states \cite{CPAst1, CPAst2, CPAst3}.

Implementing two-channel CPA requires simultaneous destructive interference between forward-scattering (transmission) and backward-scattering (reflection) fields on both sides of the medium.  
This strict realization condition may result in a single narrow CPA peak \cite{CPAla1, CPAla2}. 
To broaden the CPA peak, one can reduce the thickness of the medium to the operating wavelength or use media with stronger loss \cite{82, 83, CPAnon3}. 
These methods, however, may weaken the light transmission, and it is also challenging to suppress the reflection by space-varying refractive index. 
Moreover, many applications of CPA, such as solar energy harvesting and anti-reflecting stealth technologies, have been limited due to the narrow operable wavelength range \cite{CPAsingle}. Although broadband CPA has been explored using engineered material structures \cite{CPAbroad1, CPAbroad3, CPAbroad4} or cavity quantum electrodynamics systems \cite{CPAcav2, whitecav, whitecav02}, developing more alternative platforms for achieving two-channel broadband CPA would be important and tempting.

This paper proposes a new approach for implementing CPA using a two-channel nonreciprocal-scattering scheme with a coherent cold atomic medium subject to spatial KK modulation. On one side of the medium, the photon flow is sensitive to the relative phase of the two incident waves (of opposite directions), while on the other side, it remains unaffected. Our model exhibiting nonreciprocal scattering can serve as a four-port device (with two input and output ports) that allows for controlling the outgoing photon flow on demand \cite{CPAnon1}. Furthermore, by employing a \emph{unidirectional modulation scheme}, we simplify the strict CPA condition, which requires complete destructive interference on both sides, to a looser one related only to one-side interference. In addition, based on the chiral phase modulation and coherent absorption, we also can realize the coherent asymmetric absorber, which may have applications related to sensing \cite{CPAasy}
One can thus realize unconventional CPA with a relatively broad bandwidth. Such a broadband-perfect absorber with strong tunability has yet to be demonstrated using a coherent atomic ensemble.

\section{Model and Equations}

\begin{figure}[ptb]
\centering
\includegraphics[width=0.46\textwidth]{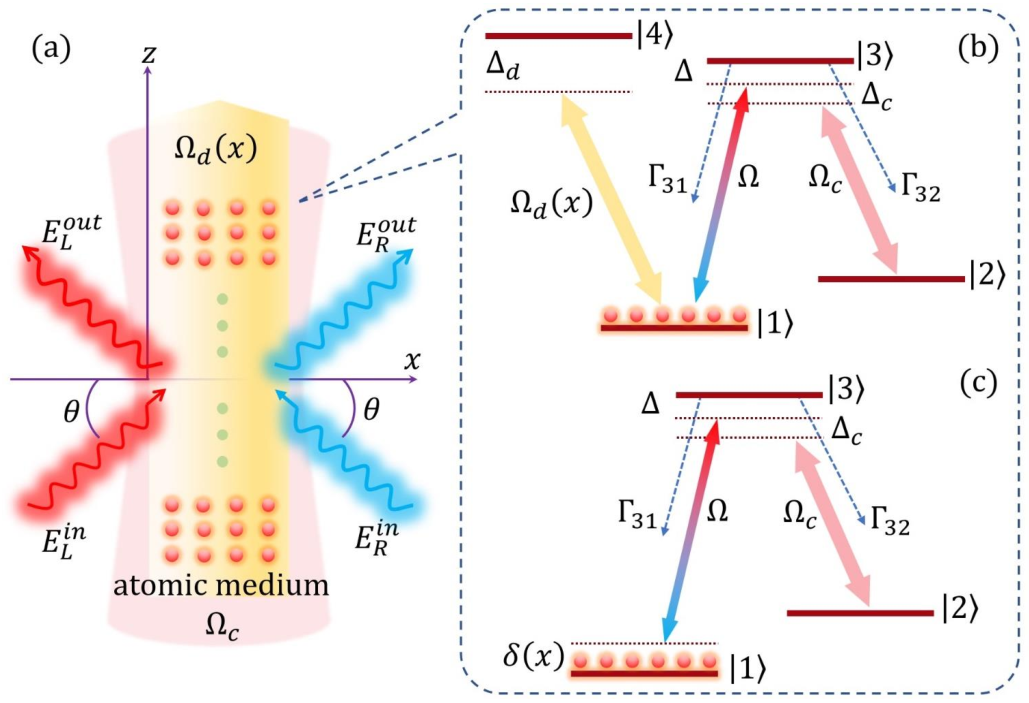}
\caption{(a) The sketch of a cold $^{87}$Rb atomic medium coupled by a strong coupling (driving) beam of Rabi frequency $\Omega_c$ ($\Omega_d$) and frequency $\omega_c$ ($\omega_d$) along the $z$-axis.
The medium is bilaterally illuminated by two weak incident waves $ \textbf{E}^{in}_L $ and $ \textbf{E}^{in}_R $, respectively, with frequency $\omega$ relative to the $\pm x$-axis at an angle $\theta$. 
(b) The four-level configuration of the $^{87}$Rb atom coupled by a coupling (driving) beam ${\Omega _c}$ [${\Omega _d(x)}$], and probed by the weak wave field ${\Omega}$. 
(c) An equivalent three-level configuration with a dynamic shift $\delta(x)$ of level $\left| 1 \right\rangle $ upon the adiabatic elimination of level $\left| 4 \right\rangle $. The population decay rate from level $\left| i \right\rangle $ to level $\left| j \right\rangle $ is denoted by ${\Gamma _{ij}}$.}\label{fig1}
\end{figure}

As shown in Fig.~\ref{fig1}(a), we consider a designed cold $^{87}$Rb atomic medium with length $L$, where each atom is driven into a four-level $N$-type configuration consisting of two excited states, $\left| 4 \right\rangle $ and $\left| 3 \right\rangle $, and two ground states, $\left| 2 \right\rangle $ and $\left| 1 \right\rangle$ as depicted in Fig.\ref{fig1}(b). These states correspond to the hyperfine levels of the $^{87}Rb$ atom, where $ \left| 4 \right\rangle =\left| 5P_{3/2} ,F = 3 \right\rangle$, $ \left| 3 \right\rangle =\left| 5P_{1/2} ,F = 1 \right\rangle$, $ \left| 2 \right\rangle =\left| 5S_{1/2} ,F = 2 \right\rangle$, and $|1 \rangle = | 5S_{1/2},F = 1\rangle$. 
A strong coupling (driving) laser beam of Rabi frequency $\Omega_c= \textbf{E}_c \cdot \textbf{d}_{23}/ 2 \hbar$ ($\Omega_d= \textbf{E}_d \cdot \textbf{d}_{14}/ 2 \hbar$) and detuning ${\Delta_c}={\omega_c}-{\omega_{32}}$ (${\Delta_d}={\omega_d}-{\omega_{41}}$) is coupled to the transition $\left| 2 \right\rangle  \leftrightarrow \left| 3 \right\rangle $ ($\left| 1 \right\rangle  \leftrightarrow \left| 4 \right\rangle $), with ${{\bf{d}}_{ij}}$ and ${\omega _{ij}}$ being the electric-dipole moment and the resonant frequency of the corresponding atomic transition. In addition, a weak resultant wave filed of Rabi frequency $\Omega= \textbf{E} \cdot \textbf{d}_{13}/ 2 \hbar$ (with $ \textbf{E} = \textbf{E}^{in}_L +\textbf{E}^{in}_R$) and detuning ${\Delta }={\omega }-{\omega _{31}}$ probes the transition $\left| 1 \right\rangle  \leftrightarrow \left| 3 \right\rangle$.

For spatial KK modulation upon the medium, the intensity of field $\Omega_c$ should be linearly varied along the $x$-axis (this can be achieved by employing a neutral density filter), i.e., $\left|\Omega _d(x) \right|^2=\left|\Omega _{d0} \right|^2x/L$ with ${\Omega _{{\rm{d}}0}}$ being the maximum at $x=L$. Using the electric-dipole and rotating-wave approximations, the interaction Hamiltonian of the four-level system can be expressed as 
\begin{align}
{H_I} =  - \hbar \left[
\begin{matrix}
0 & 0 & {\Omega  ^*} & {\Omega _d^*(x)} \\
 0 & {\Delta  } - {\Delta _c} & {\Omega _c^*} & 0 \\
 {\Omega  } & {\Omega _c} & {\Delta  } & 0 \\
{\Omega _d(x)} & 0 & 0 & {\Delta _d},
 \end{matrix} \right ].
\end{align}
Considering that field $\Omega_d$ is far-detuned from transition $|1\rangle  \leftrightarrow |4\rangle $ (${\Delta _d} \gg \Omega _{d0}, \Gamma _{41,42}$), we can adiabatically eliminate state $\left| 4 \right\rangle $. Thus, the four-level configuration reduces into an effective three-level $\Lambda$-type one, with a space-dependent energy shift $\delta(x) = | \Omega_d (x) | ^2 / \Delta_d = \delta_0 x/L$  with $\delta_{0} = | \Omega_{d0}|^2 / \Delta_d$ for level $|1\rangle$ [see Fig.~\ref{fig1}(c)].

In the weak probe limit, the complex probe susceptibility of the effective three-level system is written as
 \begin{align}
\chi \left( \Delta, x \right) = \frac{{{N_0}{{\left| {{d_{13}}} \right|}^2}}}{{2{\varepsilon _0}\hbar }}\frac{i}{{{\gamma _{31}} - i\Delta  '\left( x \right) + \frac{{{{\left| {{\Omega _c}} \right|}^2}}}{{{\gamma _{21}} + i \Delta' _c }}}},\label{eq2}
\end{align}
with $N_0$ being the atomic number density, ${\gamma _{21}} $ (${\gamma _{31}} $) representing the dephasing rate of the spin (optical) coherence, and $\Delta  '\left( x \right) = {\Delta  } + \delta \left( x \right)$ denoting the space-dependent effective detuning.
Here $ \Delta' _c =\Delta _c - \Delta '\left( x \right)$. 
The imaginary part Im[$\chi$] and real part Re[$\chi$] of $\chi$, which vary in space, govern the local absorption and dispersion properties, respectively. 
Im[$\chi$] and Re[$\chi$] can be mapped from the frequency domain to the space domain while maintaining the KK relation. 
The corresponding space-dependent refractive index can be expressed as $n=\sqrt{1+\chi}$.
The spatial KK relation can be defined by $\text{Re}[\chi(\Delta,x)]=\frac{1}{\pi}\mathcal{P}\int_{0}^{L}\frac{\text{Im}[\chi(\Delta,s)]}{s-x}ds$ \cite{KK1,KK6} with $\mathcal{P}$ the Cauthy integral principal value. 
Here, we use a well-designed field to modulate the medium instead of reconstructing the system's structure for the spatial KK relation. 
As a result, our scheme has the edge over conventional ones due to its greater tunability \cite{KK6}. 
In addition, in comparison to Ref. \cite{KK6}, we extend the spatial-KK-modulation atomic configuration to a three-level one. This extension introduces a richer set of controllable parameters and the electromagnetically induced transparency (EIT) effect, which plays a crucial role in achieving satisfactory results in this work.

The dynamical propagation of the forward component $E^+$ (backward component $E^-$) of the incident field through the $j$th layer of the medium is described by a $2 \times 2$ unimodular transfer matrix ${M_j}\left({\Delta}, d, \theta\right)$ \cite{EPBG06}, i.e., 
 \begin{align}
\left( {\begin{array}{*{20}{c}}
{E ^ + \left( {x + d} \right)}\\
{E ^ - \left( {x + d} \right)}
\end{array}} \right) = {M_j}\left( {{\Delta  },d,\theta } \right)\left( {\begin{array}{*{20}{c}}
{E ^ + \left( x \right)}\\
{E ^ - \left( x \right)}
\end{array}} \right),\label{eq3}
\end{align}
with
\begin{widetext}
\begin{equation}
{M_j}\left( {{\Delta  },d,\theta } \right) = \frac{1}{{4n'\cos \theta }}\left( {\begin{array}{*{20}{c}}
{{\left( {n' + \cos \theta } \right)}{e^{iKdn'}} - {{\left( {n' - \cos \theta } \right)}^2}{e^{ - iKdn'}}}&{\left( {{n'^2} - {{\cos }^2}\theta } \right){e^{iKdn'}} - \left( {{n'^2} - {{\cos }^2}\theta } \right){e^{ - iKdn'}}}\\
{\left( {{n'^2} - {{\cos }^2}\theta } \right){e^{iKdn'}} - \left( {{n'^2} - {{\cos }^2}\theta } \right){e^{ - iKdn'}}}&{{{\left( {n' + \cos \theta } \right)}^2}{e^{ - iKdn'}} - {{\left( {n' - \cos \theta } \right)}^2}{e^{iKdn'}}}
\end{array}} \right),\label{eq4}
\end{equation}
\end{widetext}
Here $n' = {n_j}\cos \beta $ in accordance with the law of refraction, with $\beta $ being the refraction angle and $n_j$ being the refractive index in the $j$th layer. 
The total transfer matrix $M\left( {{\Delta }, L,\theta } \right)$ of the entire medium can be obtained by multiplying transfer matrices of all layers successively. By illuminating only the left or right side of the medium, one can obtain the complex reflection and transmission coefficients for the left- and right-incident waves as
\begin{equation}
\begin{split}
\Tilde{r}_l&=r_{l} e^{i \phi _l}= -\frac{{{M_{21}}\left( {{\Delta  },L,\theta } \right)}}{{{M_{22}}\left( {{\Delta  },L,\theta } \right)}},\\
\Tilde{r}_r&=r_{r} e^{i \phi _r}= \frac{{{M_{12}}\left( {{\Delta  },L,\theta } \right)}}{{{M_{22}}\left( {{\Delta  },L,\theta } \right)}},\\
\Tilde{t}_{l,r}&=t e^{i \phi _t}= \frac{1}{{{M_{22}}\left( {{\Delta  },L,\theta } \right)}},
\end{split}\label{eq5}
\end{equation}
with the corresponding moduli $r_l,r_r,t$ and phases $\phi _l,\phi _r,\phi _t$. 

\begin{figure}[ht]
\centering
\includegraphics[width=0.48\textwidth]{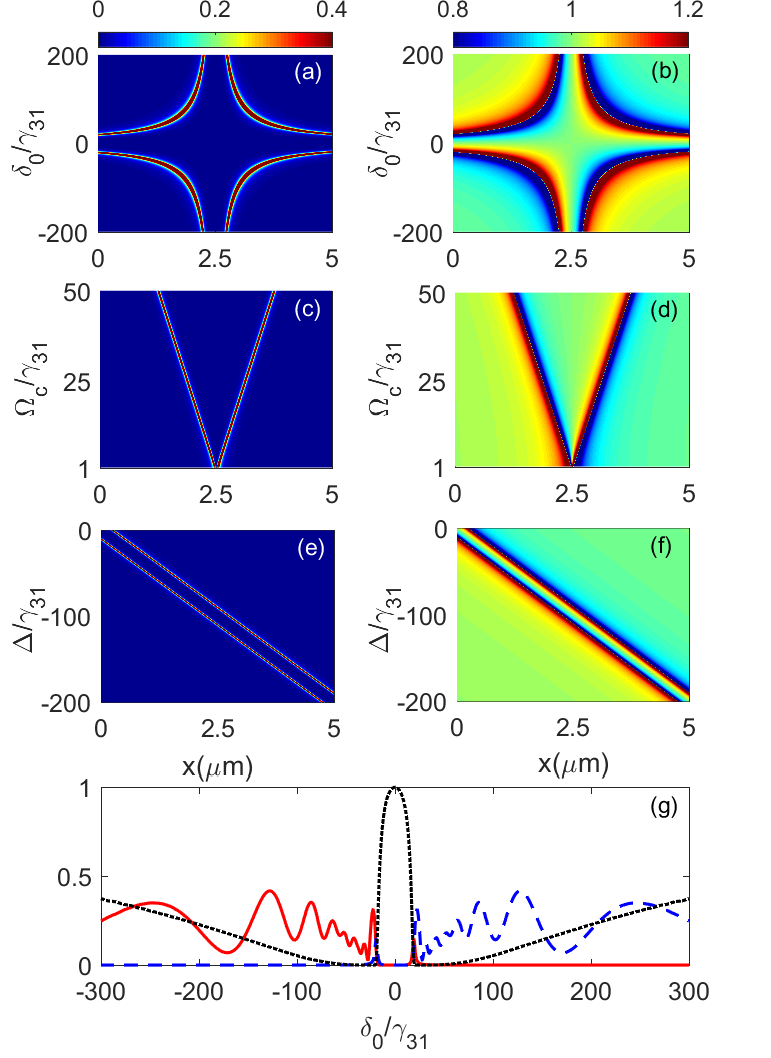}
\caption{(a) Im$[n]$ and (b) Re$[n]$ as functions of position $x$ and shift ${\delta _{\rm{0}}}$ with $\Delta =-\delta_0/2$ and ${\Omega _c} = 10{\gamma _{31}}$.
(c) Im$[n]$ and (d) Re$[n]$ as functions of position $x$ and coupling Rabi frequency ${\Omega _c}$ with $\Delta =-\delta_0/2$ and ${\delta _0} = 200{\gamma _{31}}$; (e) Im$[n]$ and (f) Re$[n]$ versus position $x$ and probe detuning $\Delta $ with ${\Omega _c} =10{\gamma _{31}}, {\delta _0} = 200{\gamma _{31}}$.
(g) Transmission amplitude $t$ (black-dotted), reflection amplitudes $r_l$ (red-solid) and $r_r$ (blue-dashed) versus shift $\delta_0$ with $\Delta =-\delta_0/2$, $\Omega _c = 10\gamma _{31}$.
Other parameters used in calculations are $\theta=0$, ${\gamma _{31}} =2\pi \times 2.87$ MHz, ${\gamma _{21}} =2\pi \times 1.0 $ KHz, ${d_{13}} = 1.79 \times {10^{ - 29}} $ C$ \cdot $m, ${N_0} = 2.0 \times {10^{13}}$ cm$^{-3}$, $L = 5.0$ $\mu$m, ${\Delta _c} = 0$ and ${\lambda  } = 795$ nm.}\label{fig2}

\end{figure}

By using Eqs.~(\ref{eq2}) and (\ref{eq5}), one can investigate the optical response of a tunable atomic medium subjected to spatial KK modulation. 
Figure~\ref{fig2} shows that the complex refractive index $n$ can be displaced along the $x$-axis with the incident detuning $\Delta$, coupling Rabi frequency $\Omega_c$, and level shift $\delta_0$ (related to the intensity and frequency of the driving beam), respectively. 
In the spatial domain, $n$ exhibits a KK profile (with an even-like Re$[n]$ and an odd-like Im$[n]$) and two peaks due to the equivalent three-level structure. 
As shown in Figs.~\ref{fig2}(a) and (b), when $|\delta_0|$ approaches $0$, the double peaks move to opposite directions, respectively, and the profiles of $n$ gradually broaden. 
The peaks disappear at $\delta_0=0$ due to $n=1$ on resonance, corresponding to the regime of EIT \cite{EIT01}, where, however, the optical response is space-independent. 
Figures~\ref{fig2}(c) and (d) demonstrate the linear increase in separation between double peaks with ${\Omega_c}$. 
However, when $\Delta$ varies, the peaks move linearly along the medium together due to the equivalent contribution of $\Delta $ and $\delta(x)$ to $n$ in Eq.(\ref{eq2}). 
Under spatial KK modulation, the medium exhibits a unidirectional reflectionless feature, as shown in Fig.\ref{fig2}(g). 
When $|\delta_0|<20\gamma_{31}$, the KK profile of $n$ is much broader than the medium, rendering $n$ approximately uniform throughout the medium, thereby breaking the spatial KK relation. 
Consequently, the EIT window appears, and the reflection is suppressed. 
With an increase in $|\delta _0|$ (leaving the EIT window), the transmission significantly decreases, and reflection appears. 
Note that $r_r=0$ ($r_l=0$) when $r_l\neq0$ ($r_r\neq0$), i.e., unidirectional reflectionlessness is realized in the regime of spatial KK modulation. 
Moreover, the reversal of the KK profiles of $n$ [see Figs.~\ref{fig2}(a) and (b)], via changing $\delta _{0} \to -\delta _{0}$, enables the switching of the side of unidirectional reflectionlessness.

To observe two-channel scattering, we employ the scattering matrix $S$ that can relate the incident wave amplitudes $E_{R, L}^{in}$ to the outgoing ones $E_{L, R}^{out}$ [see Fig.~\ref{fig1}(a)], as
 \begin{align}
\left( {\begin{array}{*{20}{c}} {E_L^{out}}\\
{E_R^{out}} \end{array}} \right) = S\left( {\begin{array}{*{20}{c}} {E_R^{in}},\\
{E_L^{in}{e^{i\phi }}}\end{array}} \right) = \left( {\begin{array}{*{20}{c}} {{\Tilde{t}_r}}&{{\Tilde{r}_l}}\\
{{\Tilde{r}_r}}&{{\Tilde{t}_l}}
\end{array}} \right)\left( {\begin{array}{*{20}{c}}
{E_R^{in}},\\
{E_L^{in}{e^{i\phi }}}
\end{array}} \right),\label{eq6}
\end{align}
with $\phi$ being the relative phase between the two incident waves. 
Then, we obtain
\begin{equation}
\begin{split}
&E_L^{out} = t{e^{i{\phi _t}}}E_R^{in} + {r_l}E_L^{in}{e^{i\left( {{\phi _l} + \phi } \right)}},\\
&E_R^{out} = {r_r}{e^{i{\phi _r}}}E_R^{in} + tE_L^{in}{e^{i\left( {{\phi _t} + \phi } \right)}}.  \label{eq7}
\end{split}
\end{equation}
The scattering intensities on both sides are
\begin{subequations}
\begin{align}
&{S_L} = {\left( {{r_l}E_L^{in}} \right)^2} + {\left( {tE_R^{in}} \right)^2} + 2{r_l}t E_L^{in} E_R^{in}\cos \left( { \phi + {\phi _l}- {\phi _t}} \right),\label{eq8a}\\
&{S_R} = {\left( {{r_r}E_R^{in}} \right)^2} + {\left( {tE_L^{in}} \right)^2} + 2{r_r}t E_L^{in} E_R^{in}\cos \left( {  \phi+{\phi _t}  - {\phi _r}} \right).\label{eq8b}
\end{align}
\end{subequations}
Each scattering intensity comprises transmission and reflection components and an interference term involving phases. 

If one requires the absence of the left-side scattering (${S_L} = 0$), the coherent condition must be satisfied as
\begin{equation}
    \begin{split}
&{\phi _l} + \phi={\phi _t} + \left( {2k + 1} \right)\pi,\\
&t\sqrt {{I_R}}={r_l}\sqrt {{I_L}},
\label{eq9}
    \end{split}
\end{equation}
with $k = 0, \pm 1, \pm 2 \cdots$. Here ${I_i}= \frac{1}{2}\varepsilon_0 c E_i^{in}{\left( {E_i^{in}} \right)^ * }$ ($i=R,L$) represents the intensity of incident waves.
Likewise, ${S_R} = 0$ should satisfy the coherent condition as
\begin{equation}
    \begin{split}
&{\phi_t}+\phi={\phi _r}+\left({2k + 1}\right)\pi,\\
&t\sqrt{{I_L}}={r_r}\sqrt {{I_R}}.
\label{eq10}
    \end{split}
\end{equation}
Then CPA can be achieved by satisfying both Eqs.~(\ref{eq9}) and (\ref{eq10}), which represent the co-occurring complete destructive interference on both sides.

\section{Spatial KK modulation induced chiral phase modulation}

\begin{figure}[tp]
\centering
\includegraphics[width=0.47\textwidth]{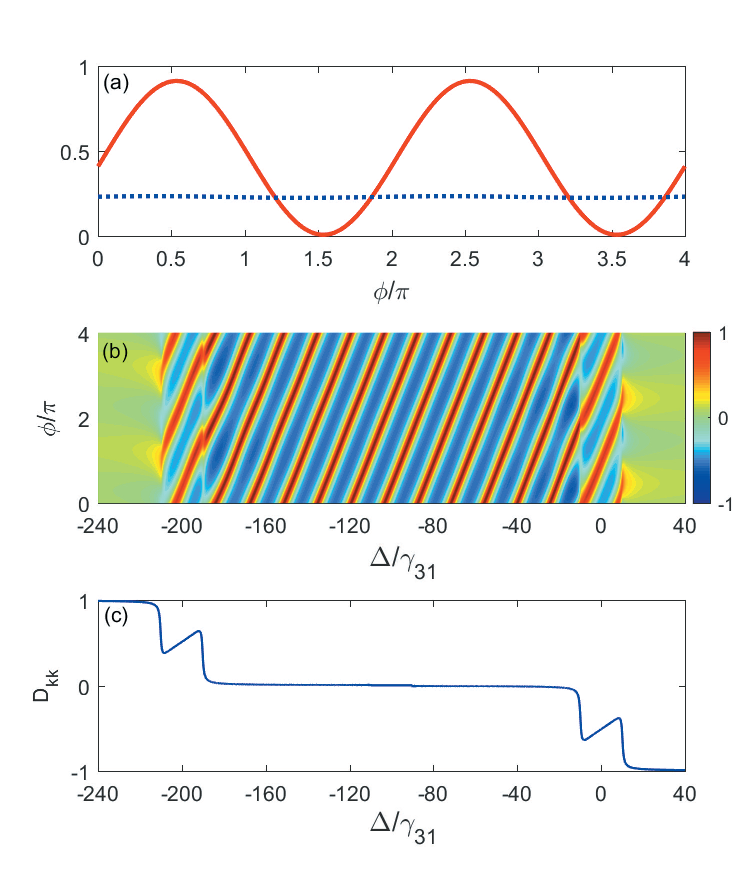}
\caption{(a) Right-side scattering ${S_R}$ (red-solid) and left-side scattering ${S_L}$ (blue-dotted) versus the relative phase $\phi$ with $\Delta =-100\gamma_{31}$.
(b) Contrast $C$ as a function of detuning $\Delta $ and relative phase $\phi$.
(c) Figure of merit $D_{kk}$ verse detuning $\Delta$.
Here $\delta_0=200\gamma_{31}$, $I_R=1.15$, $I_L=1$, $\Omega_c = 10\gamma _{31}$, and other parameters are the same as in Fig.~\ref{fig2}.}
\label{fig3}
\end{figure}

The two-channel scattering is sensitive to the incident phase and is influenced by the space-dependent complex refractive index.
Consequently, under spatial KK modulation, it will significantly differ from conventional scattering. 
To confirm the parameter regime for spatial KK modulation, we use the following integral as a figure of merit for the spatial KK relation:
\begin{equation}
D_{kk}(\Delta)=\frac{\int_{0}^{L}\left\{\text{Re}[\chi(\Delta,x)] -\frac{1}{\pi}\text{P}\int_{0}^{L}\frac{\text{Im}[\chi(\Delta,s)]}
{s-x}ds\right\} dx}{\left\vert \int_{0}^{L}\text{Re}[\chi(\Delta,x)dx\right\vert }. \label{Eq11}
\end{equation}
Here, $D_{kk}=0$ corresponds to the unbroken KK phase, where the spatial KK relation is fully satisfied; $\left| D_{kk} \right|=\pm1$ denotes the broken KK phase, where the relation is entirely destroyed; the transition regime between these two phases is characterized by as $1<\left| D_{kk} \right|<0$.

For the analysis, we will use the relative intensities of incident waves $I_{R, L}$ and scattering $S_{R, L}$, which are described in terms of the unit of $ \frac{1}{2}\varepsilon_0 c|\textbf{E}_0|^2$ and under the regime of the weak field limit ($\Omega_0= \textbf{E}_0 \cdot \textbf{d}_{13}/ 2 \hbar\ll \Omega_{c,d}, \gamma_{31}$). 
Figure~\ref{fig3}(a) displays the nonreciprocal behavior of the right- and left-side scattering $S_{R, L}$ against the relative phase $\phi$ in the regime of spatial KK modulation. 
Notably, $S_L$ is immune to the phase, while $S_R$ exhibits typical periodical variation. 
This is caused by the fact that $S_L$ losses the interference term and solely consists of the non-zero transmission component due to the negligible $r_l$ arising from the one-way suppression of reflection under spatial KK modulation, coinciding with Eq.~(\ref{eq8a}).  
In other words, we achieve one-side interference for two-channel scattering arising from spatial KK modulation. 
Thus, a noteworthy feature in phase sensitivity is its dependence on direction, which could be referred to as the chiral (unidirectional) phase modulation upon scattering behavior, whose underlying physics is that the complex coupling between the atomic medium and light waves depends on the direction. 
This may have the potential to enhance the burgeoning field of chiral quantum optics \cite{unT7, cqo01, cqo02, cqo03}. 
 
The chiral phase modulation enables the nonreciprocal scattering typically, with a few exceptions where it reverts to reciprocity, as shown in Fig.~\ref{fig3}(a). 
Unidirectional (perfect nonreciprocal) scattering occurs at certain phase points, related to the one-side complete destructive interference. 
This can enable the coherent asymmetric absorber.
It allows for stable outflows on one side while absorbing them completely on the other. 
In typical four-port devices, the signals from two output ports are phase-sensitive \cite{CPA02, CPA03}. 
However, our scheme offers the possibility of the phase modulation on one-side photon flows through chiral phase modulation while keeping the other unaffected. 
This feature could be exploited to develop a unidirectional transmission modulator and an all-optical unidirectional switch \cite{CPAnon1, asy01, asy02}. 

To demonstrate whether scattering is reciprocal or nonreciprocal, we plot the scattering contrast $C$ for $S_{L, R}$ in Fig.~\ref{fig3}(b), given by 
\begin{equation}
 C =\left\vert\frac{S_L-S_R}{S_L+S_R}\right\vert. \label{eq12}
\end{equation} 
As shown in Fig.~\ref{fig3}(c), $C$ also exhibits periodic changes between $0$ (reciprocity) and $1$ (unidirectionality) with $\phi$ in the range of $\Delta \in[-190\gamma_{31},-10\gamma_{31}]$, corresponding to the unbroken KK phase where $|D_{kk}|=0$. 
However, achieving perfect nonreciprocity is not possible in the regions of $[-210\gamma_{31},-188\gamma_{31}$] and $[-12\gamma_{31},10\gamma_{31}]$, which correspond to the phase transition region where $0<|D_{kk}|<1$. 
The scattering behavior exhibits trivial reciprocity when modulating the medium into the broken KK phase. 
Furthermore, the nonreciprocity also depends on the incident wave frequency. 
We note that reversing the direction of unidirectional scattering is possible by changing the sign of $\Delta_d$. 
This scheme allows for the unidirectional flow of light waves of specific frequencies by adjusting the phase, and the direction of unidirectionality can also be switched without changing the incident wave intensity. 
This feature can be used to develop photonic filters based on phase modulation. 
Note that the relative phase can be expressed as $\phi = k\Delta L$ with $k$ being the wavevector and $\Delta L$ optical path difference that can be adjusted by introducing an additional delay on the path of one of the light waves in experiments. 
Thus, this phase modulation on scattering is a linear operation and results in reduced power requirements. 

\section{Other tunability on nonreciprocal scattering}

\begin{figure}[t]
\centering
\includegraphics[width=0.5\textwidth]{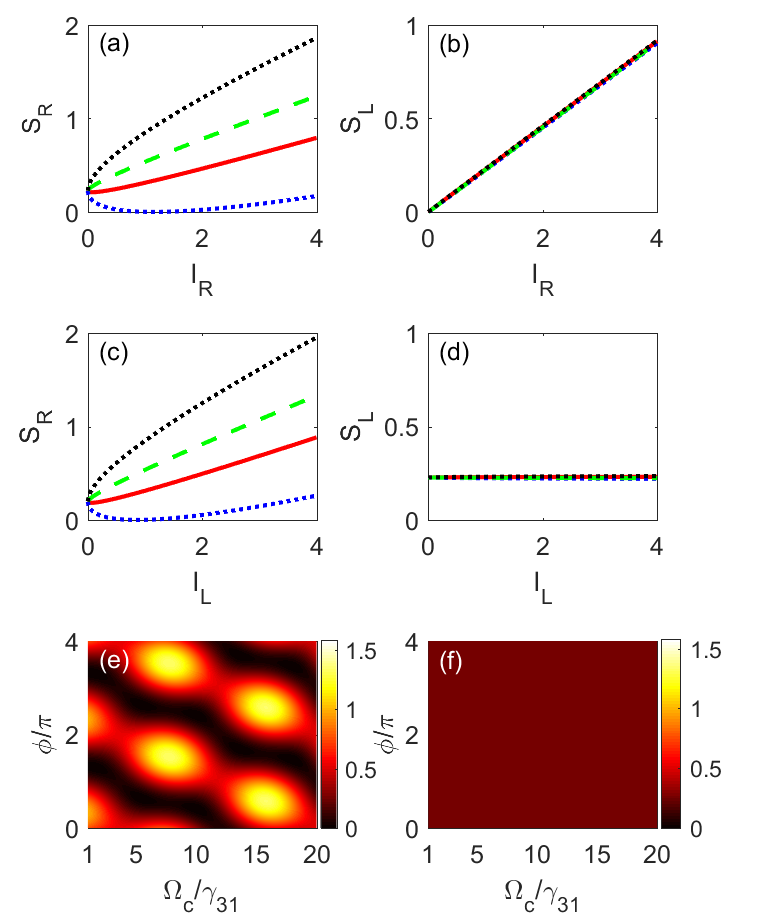}
\caption{(a) Right-side scattering $S_R$ and (b) left-side scattering $S_L$ versus incident intensity $I_R$ with fixed $I_L=1$.
(c) $S_R$ and (d) $S_L$ versus $I_L$ with fixed $I_R=1$. Curves represent $\phi=0$ (red-solid), $\phi=0.4\pi$ (blue-dotted), $\phi=\pi$ (green-dashed), $\phi=1.4\pi$ (black-dot-dashed).
(e) $S_R$ and (f) $S_L$ as functions of coupling Rabi frequency $\Omega _c$ and relative phase $\phi$ with $I_R=1.15$ and $I_L=1$. Here $\delta_0=200\gamma_{31}$ and $\Delta =-\delta_0/2$, and other parameters are the same as in Fig.~\ref{fig2}.}\label{fig4}
\end{figure}

We explore other tunable methods for manipulating nonreciprocal scattering, including the effect of incident wave (coupling beam) intensity. 
Figures~\ref{fig4}(a) and (c) exhibit similar nonlinear variations in the right-side scattering $S_R$ as one of the two incident waves is increased, provided when $r_r$ and $t$ are relatively close corresponding to Eq.~(\ref{eq8b}). 
This characteristic is commonly observed in other studies about controlling light with light, such as with planar metal metamaterials \cite{CPAnon1}. 
However, in contrast, nontrivial behavior of scattering on the phase-non-sensitivity side will occur with the incident intensities, as shown in Fig.~\ref{fig4}(b) and (d), respectively. Specifically, $S_L$ is proportional to $I_R$ but remains unchanged with $I_L$. 
This observation can be elucidated by Eqs.~(\ref{eq8a}) and (\ref{eq8b}) and is attributed to the chiral phase modulation. 
This asymmetric controlling light with light is of interest for asymmetrical tasks in quantum communication. 

The effects of coupling beam intensity on nonreciprocal scattering are shown in Fig.~\ref{fig4}(e) and (f). 
Scattering $S_R$ on the phase-sensitivity side can vary periodically with the coupling intensity and phase. 
However, $S_L$ on the phase-non-sensitivity side remains robust against the coupling intensity and phase. 
The maximum value of $S_R$ can reach $1.28$ when $I_L=1$ and $I_R=1.15$ via the competition between constructive interference and atomic decay, resulting in the amplification of the single-sided output optical signal. 
This coherent atomic scheme, as a four-port device, may serve as a tunable all-optical amplifier compared to other metamaterials \cite{CPAnon1}. 
Notably, this mechanism has the advantages of not requiring additional gain pumping, having no harmonic noise, and possessing intrinsic nonlinearity \cite{CPArevline}.

\section{Broadband CPA platform}

\begin{figure}[ptb]
\centering
\includegraphics[width=0.47\textwidth]{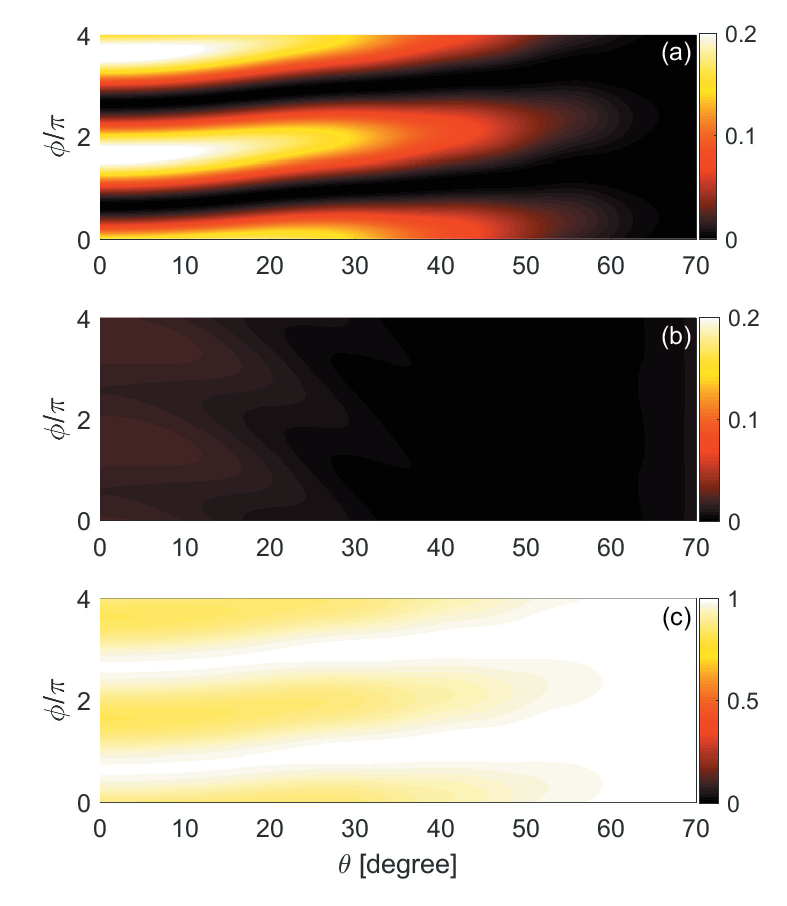}
\caption{(a) Right-side scattering $S_R$, (b) left-side scattering $S_L$, and (c) absorption fact $A$ as functions of incident angle $\theta$ and relative phase $\phi$ with ${\delta _0} = 100{\gamma _{31}}$ and $\Delta =-\delta_0/2$. The intensity ratio of two incident waves $I_R/I_L$ satisfies Eq.~(\ref{eq10}) with $I_L=1$. Other parameters are the same as in Fig.~\ref{fig2}.}
\label{fig5}
\end{figure}

By exploiting chiral phase modulation, we investigate the scattering behavior of coherent atomic media over a wide range of incident angles $\theta$ for a perfect absorber. 
As $\theta$ increases, the propagating path in the medium lengthens, accounting for contact with more atoms and then increasing atomic absorption. 
Using an atomic ensemble, the perfect absorber may be achieved through stronger medium absorption and wave interference. 
To quantify absorption performance with energy conservation, the absorption factor is used as
\begin{equation}
A=\frac{(I_L+I_R-S_L-S_R)}{(I_L+I_R) }. \label{eq13}
\end{equation}

Figure~\ref{fig5}(a) shows that as $\theta$ increases on both sides, $S_R$ varies periodically between the maxima $0.22$ and the minima $0$ with $\phi$ due to interference. 
For relatively large values of $\theta$, the component of the incident wave that could enter the medium is so weak that the scattering can be ignored. 
In Fig. 5(b), the disappearance of $S_L$ occurs because the originally weak transmission of the right-incident wave [see Fig.~\ref{fig2}(g)] is perfectly absorbed as $\theta$ is increased. 
Thus, with appropriate values of $\phi$, both-side scattering can be eliminated simultaneously due to the interference and absorption, the process of which is restricted by the condition in Eq.~(\ref{eq9}), but not by Eq.~(\ref{eq10}). 
These behaviors are visible in Fig.~\ref{fig5}(c), where some bands of $A \simeq 1$ appear, indicating CPA. 
When $\theta$ becomes sufficiently large ($60^\circ \le \theta \le 90^\circ$), the incident wave may be almost entirely absorbed normally due to the spontaneous decay of atoms, which is unrelated to the coherent process. In contrast to the stringent conditions of traditional CPA, which requires satisfying both Eqs.~(\ref{eq9}) and (\ref{eq10}) simultaneously, the requirement of CPA in our coherent atomic media is significantly relaxed, as only one of those coherent conditions needs to be fulfilled. 
Then, this simplification may allow for achieving a perfect absorber with greater ease and over a broader range of spectra.

\begin{figure}[ptb]
\centering
\includegraphics[width=0.5\textwidth]{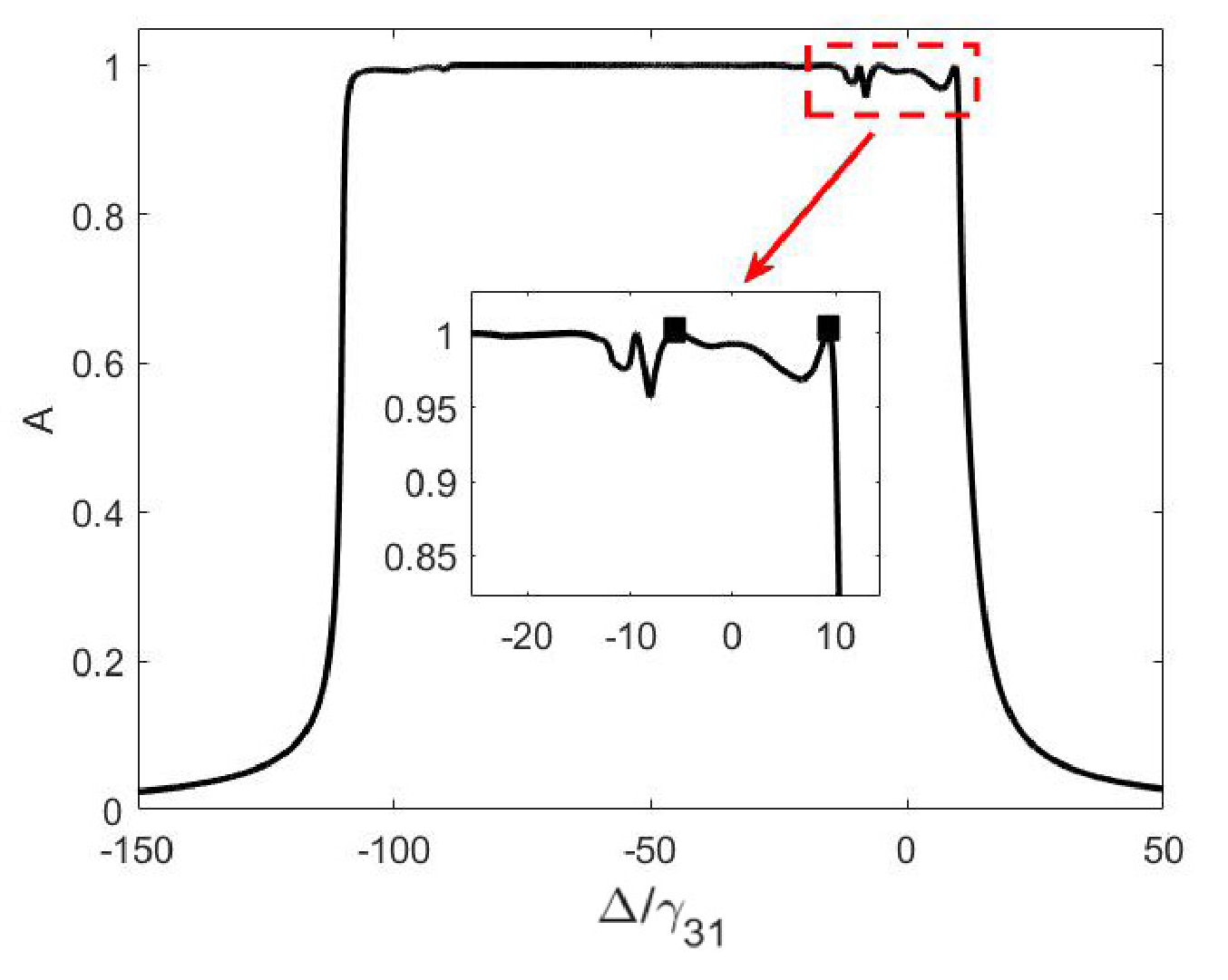}
\caption{Maxima of absorption factor $A$ as a function of detuning $\Delta$ with ${\delta _0} = 100{\gamma _{31}}$, over the incident angle $\theta$ ranging from $0^\circ$ to $60^\circ$ and the relative phase $\phi$ ranging from $0$ to $2\pi$. The intensity ratio of two incident waves $I_R/I_L$ satisfies Eq.~(\ref{eq10}) with $I_L=1$. The insert is an enlarged view. Other parameters are the same as in Fig.~\ref{fig2}.}
\label{fig6}
\end{figure}

Figure~\ref{fig6} shows the investigation of the maximum absorption over the incident angle $\theta$ ranging from $0^\circ$ to $60^\circ$ and relative phase $\phi$ from $0$ to $2\pi$. 
In the unbroken KK phase, the absorption factor $A > 0.98$ over the frequency interval $[-110\gamma_{31}, -10\gamma_{31}]$, indicating the presence of a broadband continuous CPA platform that drops off sharply beyond these intervals. 
This unique platform has a flat top and sharp edge, making it highly sensitive to incident wave frequency. 
The coherence requirement on both sides for typical CPA results in one point or only a few isolated CPA points in spectra \cite{CPA01, CPAnon3}. 
This case can be found in the range of $[-12\gamma_{31}, -10\gamma_{31}]$ (see the insert of Fig.~\ref{fig6}). 
Within the KK phase transition region, this range corresponds to asymmetric reflection rather than unidirectional reflectionlessness, and thus, the CPA points only occur when both-side CPA condition is met. 
The underlying physics for the broadband CPA is that the coherent atomic system's tunability provides sufficient tunable parameters combined with significant simplification of the coherent condition. 
This CPA platform and nonreciprocal scattering may be used for special all-optical logic gates by defining the threshold intensity \cite{CPAnon1}. 
The optical nonreciprocal behavior based on CPA also has been demonstrated in photonic lattices \cite{ucpa}. 

It is noteworthy that the coherent atomic system's unique properties may make it an excellent platform for integrating functional devices and circuits \cite{chip0} while atom chips have been developed using nanofabricated wires to trap and guide atoms \cite{chip1}. 
These atom chips have been utilized for quantum metrology, information processing, and device miniaturization and integration \cite{chip2,chip3,chip4}. 
Hence, it is reasonable to anticipate that our scheme could be implemented on atom chips to further integrate all-optical devices for optical communication networks and photonic information processing.

\section{Conclusion and outlook}

We use a coherent $^{87}$Rb atomic ensemble to investigate the effects of spatial KK modulation on two-channel scattering, aiming to identify potential applications in nonreciprocal scattering and chiral phase modulation with abundant tunable parameters. 
The spatial-KK-relation-induced one-sided interference results in the chiral phase modulation that allows us to achieve perfect nonreciprocal scattering, enabling an asymmetric absorber and photonic filter with phase modulation. 
The scattering on the phase-sensitivity side periodically varies with the phase and coupling beam intensity and shows typical nonlinear variation with the intensity of both incident waves. 
However, the scattering on the phase-non-sensitivity side is immune to the phase and coupling intensity and robust to incident wave intensity from the same side but linearly varies with the incident wave from the other side. 
This can act as a unidirectional amplifier based on one-side constructive interference. 
With the looser condition for perfect absorption and tunability in this system, we realize a coherent perfect absorber that can operate over a wide range of incident angles and exhibit a broadband platform with sharp edges. 

Our study suggests that this coherent atomic continuous medium exhibits excellent potential for a wide range of four-port two-channel all-optical devices. 
These devices include, but are not limited to, special light coherence filters/splitters, optical switches, unidirectional modulators and amplifiers based on nonreciprocal scattering and chiral phase modulation, as well as coherent photodetectors, logical gates, photonic sensors, and variable attenuators. 
These devices can operate at extremely low power levels in the EIT regime. 
The ability to flexibly control nonreciprocal scattering properties is of expectation to expedite the development of integrated and miniaturized photon devices and communication technologies.

\section*{Acknowledgment}
This work is supported by the National Natural Science Foundation of China (12074061).

\end{document}